\documentclass{ws-ijmpa}
\pdfoutput=1 
\begin{document}

\markboth{Teresa Montaruli for the ANTARES Collaboration}
{The ANTARES underwater neutrino telescope}

%
\catchline{}{}{}{}{}
%

\title{The ANTARES underwater neutrino telescope}

\author{TERESA MONTARULI\footnote{on leave from Bari University and INFN, Physics Department, 70126, Italy}  for the ANTARES COLLABORATION \footnote{
http://antares.in2p3.fr}}

\address{Department of Physics, University of Wisconsin - Madison, 1150 University Avenue,\\
Madison, Wisconsin, 53706, USA
\\
tmontaruli@icecube.wisc.edu}

\maketitle

\begin{history}
\received{Day Month Year}
\revised{Day Month Year}
\end{history}

\begin{abstract}
ANTARES is the first undersea neutrino telescope. It is in its complete configuration since 
May 2008 at about 2.5 km below the sea surface close to Marseille. Data from 12 lines are being analyzed and are producing first results. Here we discuss first analysis results for 5 lines and 10 lines, and we also comment on the performance of the full detector. We show that the detector has  capabilities for discriminating upgoing neutrino events from the much larger amount of downgoing atmospheric muons and that data and simulation are in good agreement.  We then discuss the physics reach of the detector for what concerns point-like source and dark matter searches.
\end{abstract}

\ccode{PACS numbers: 11.25.Hf, 123.1K}

\section{Underwater experiments}	

\begin{figure}[pb]
\centerline{\psfig{file=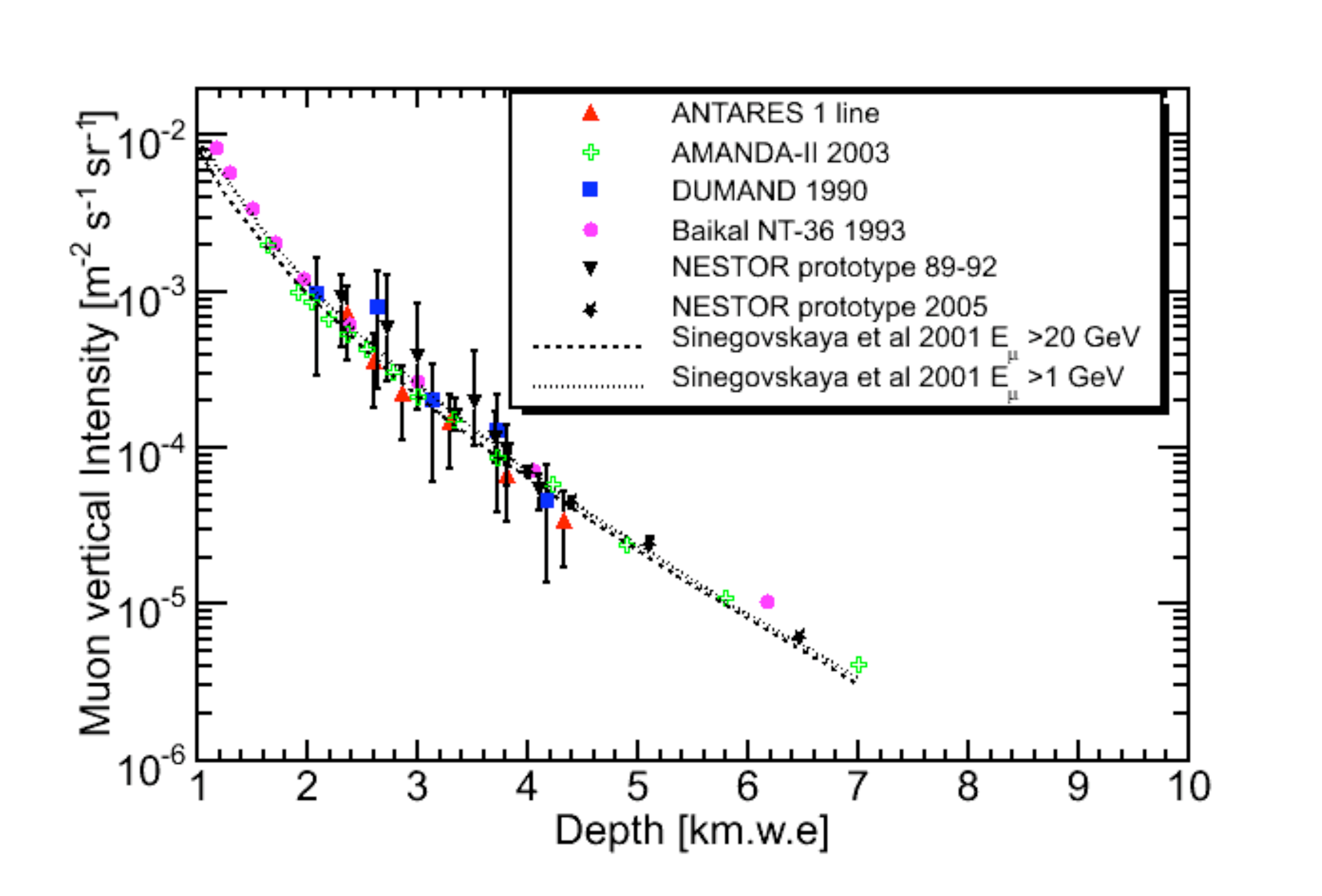, width=10cm}}
\caption{Vertical muon flux in under-water and under-ice detectors. Experiment references are in \protect\cite{dumand,baikal,dpb,nestor}. The error on ANTARES is 50\%. Predicted curves are from Ref.~\protect\cite{sine} and for visible energy of muons larger than 20 GeV and 1 GeV. 
\label{fig1}}
\end{figure}

The long standing efforts for building Cherenkov detectors in sea depths started with
the DUMAND project that measured the muon vertical intensity at depths ranging between 2 and 4 km with a prototype line at a depth of 4.8 km about 30 km off-shore the island of Hawaii in Nov. 1987 \cite{dumand}. The project was canceled in 1995, while in 1993 a first configuration of 36 phototubes (PMTs) on 3 strings (NT-36) was installed in Lake Baikal in Siberia at the shallow depth of 1.1 km, 3.6 km off-shore \cite{baikal}. The latest configuration of the experiment (NT-200+) was put into operation in Apr. 2005, with an umbrella-like structure of 8 strings 72 m long,  with pairs of up-looking and down-looking PMTs with 37-cm photocatodes. Three external strings at 100 m from the center of NT200 have been added to increase the sensitivity at very high energies for cascades. Lake water compared to sea water offers the advantage of being a quieter environment and the frozen surface in winter can be used as a platform for deploying and replacing broken components. Nonetheless, the optical background
can be more seasonal dependent than in sea water and the absorption length for blue light (480 nm) is about 20 m, while in sea water it is typically between 50-60 m. Both for sea water and lake water the scattering length, which mainly affects the angular resolution of neutrino telescopes, is between 15-70 m, but the angular distribution of scattered photons is very forward peaked ($<\cos\theta> \sim 0.85-0.9$) 
so that the effective scattering length ($L_{eff} = L_{scatt}/(1-<cos\theta>)$
can be up to 200-300 m. 
The NESTOR collaboration begun surveys of an area close to the Peloponnese coast at a mean depth of 4 km in the '90s. They measured recently the muon flux with a hexagonal floor with 12 up-looking and down-looking PMTs 
\cite{nestor} and it is studying the possibility to deploy towers of 12 of these floors. 
Fig.~\ref{fig1} summarizes the muon flux measurements of all these projects. It is noticeable how the agreement of the measurement has improved with time.  In the plot, we also show the results for the first line of ANTARES detector described below \cite{line1}.

\section{The ANTARES experiment and its working principle}	

\begin{figure}[pb]
\begin{center}
\psfig{file=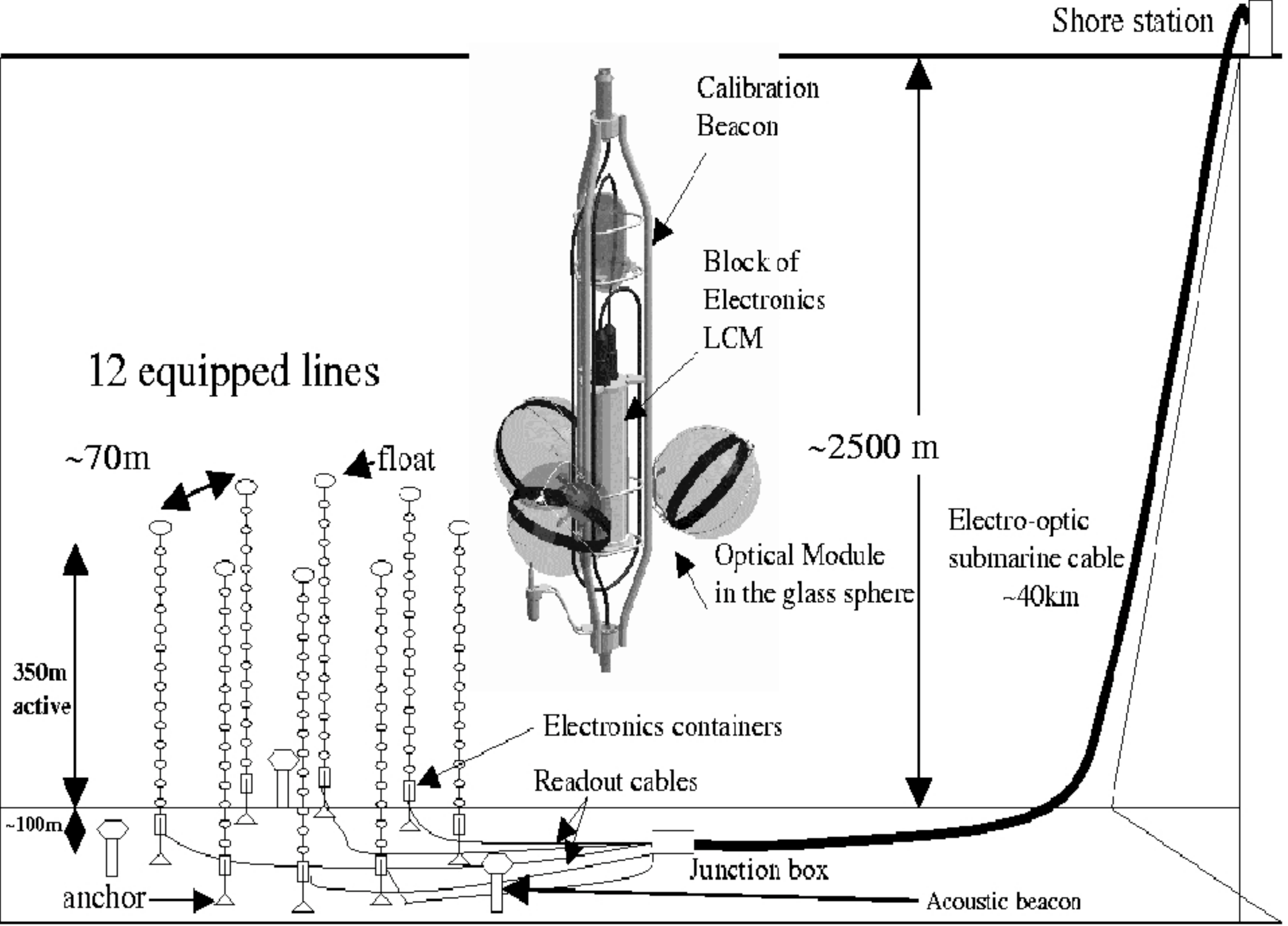, width=9cm}
\end{center}
\caption{A scheme of the 12 lines, the JB and EOC of the ANTARES detector. Also a storey is represented in an enlarged scale.  \label{fig2}}
\end{figure}

The ANTARES Collaboration, joining about 150 physicists, engineers, and sea experts from 7 European countries (France, Germany, Italy, Romania, Russia, Spain, The Netherlands) and 24  Institutions, formed in 1996. Marine properties of the ANTARES site off-shore La Seyne sur Mer (South France) were studied between 1996 and 2000 in about 45 sea campaigns. The properties of light propagation at the ANTARES site, 2475 m below the sea level, are summarized in Ref.~\cite{transmission}: the absorption length in the blue is about 60 m while the effective scattering length is about 250 m.
The milestones of the detector construction are the deployment of the 45 km-long main electro-optical cable (EOC) in 2001 and of the junction box (JB) at its end in 2002. Between 2003 and 2005 various prototype lines and a mini-instrumentation line for environmental parameter measurements where deployed  \cite{milom,line0} and in 2006 the first 2 lines of the detector. In 2007 further 8 lines where deployed and on May 30, 2008 the whole detector made of a total of 12 lines was put into operation. A scheme of the detector and of the line layout is shown in Fig.~\ref{fig2}. Lines were connected during 5 submarine operations, one conducted by a submarine and all others by an unmanned Remote Operated Vehicle. 

ANTARES is made of 12 lines held taught by buoys and anchored at the sea floor connected to the JB that distributes power and data from/to shore. The instrumented part of the line starts at 100 m above sea level so that Cherenkov light can be seen also from the region below. Lines are separated by 60-75 m from each other and each of the lines holds 25 floors called storeys. Each storey has 3 PMTs (Hamamtsu 10" \cite{pmt}) looking downward at $45^{\circ}$ from the vertical housed inside pressure resistant glass spheres made of two halves closed by applying an under-pressure of 200-300 mbar. The set-up including the PMT, the glass sphere, the silicon gel for optical coupling between the glass and the photocathode and the mu-metal cage for shielding the Earth magnetic field is called optical module (OM) \cite{om}. Storeys also include titanium containers housing the frontend electronics with a pair of ASIC chips per PMT used for signal processing and digitization that provide the time stamp and amplitude of the PMT signal. Each of the OMs contains a pulsed LED for calibration of the relative variations of PMT transit time and a system of LED and laser Optical Beacons allows the relative time calibration of different OMs. An internal clock system distributes from shore the 20 MHz clock signal, that is synchronized by GPS to the Universal Time with a precision of $\sim 100$ ns. Time calibrations allow a precision at the level of 0.5 ns ensuring the capability of achieving an angular resolution at the level of $0.3^{\circ}$ for muons above 10 TeV \cite{calibration}. ANTARES is equipped with a positioning system that includes tiltmeters and compasses giving the orientation of storeys and an acoustic triangulation system of hydrophones and transceivers that reconstruct line shapes. Measurement of line displacements show that lines move coherently in the radial direction and that the achievable precision 
on the relative position is better than 10 cm. This information can be stored in databases and used by the offline event reconstruction.

\begin{figure}[pb]
\begin{tabular}{cc}
\psfig{file=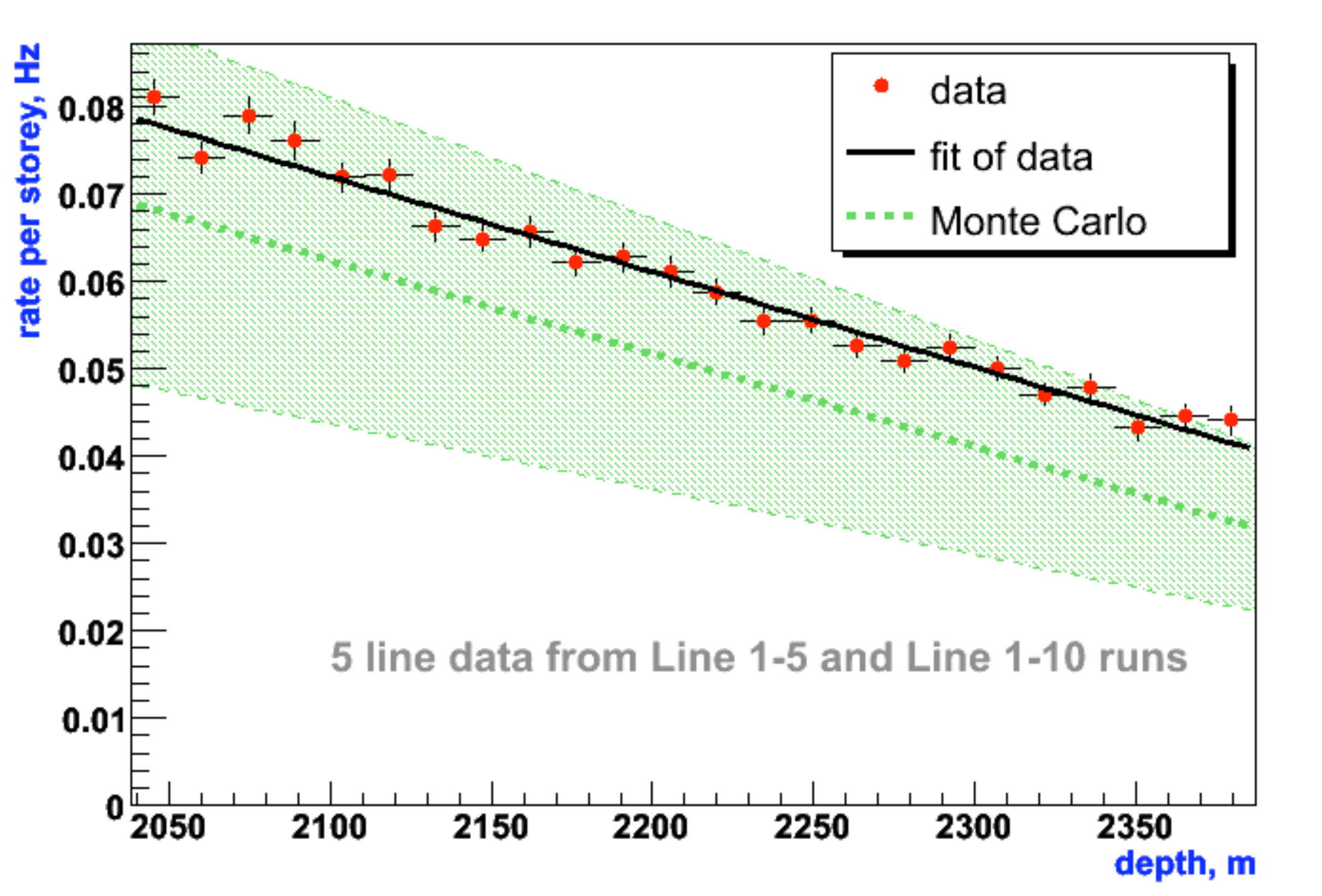, width=6.5cm,height=5.5cm}&
\psfig{file=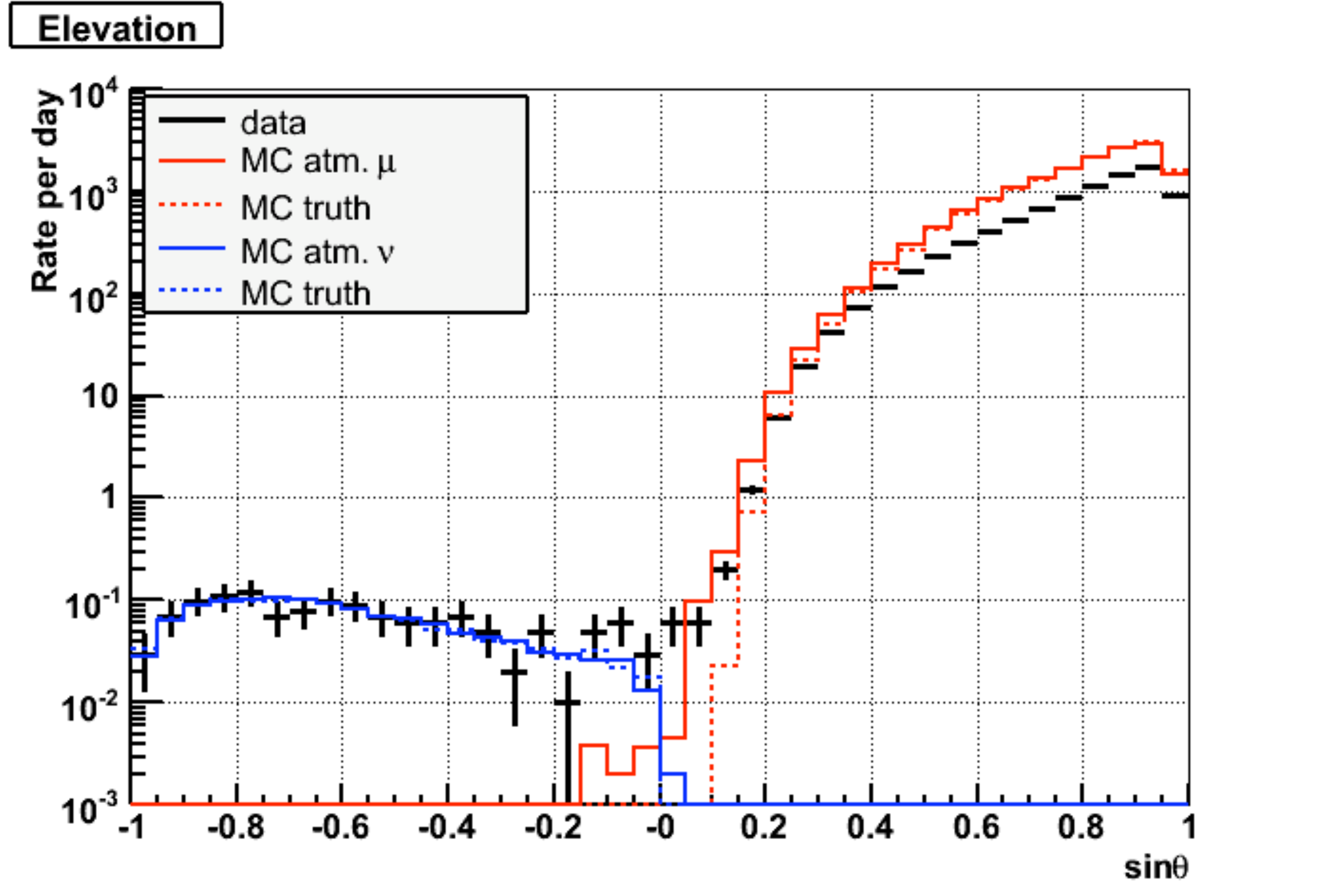, width=6.5cm,height=5.5cm}
\end{tabular}
\caption{
On the left: Rate of coincidences between 2 storeys separated by about 14.5 m due to muon tracks as a function of depth for data (red circles with statistical errors) and MC shown with an uncertainty of about 30\%. 
On the left: sine of the zenith angle distribution for 120.5d of 5 line data (with statistical errors) compared to MC (solid line for reconstructed, dashed line for true muons on the right and neutrinos on the left). \label{fig3}
}
\end{figure}

All data ($\sim 7.5$ Gb/s) are sent to shore satisfying a L0 condition requiring hits above a threshold of 1/3 of a photoelectron (pe). A L1 is formed by a local coincidence of 2 L0 in a storey in a 20 ns window or a large pulse above a few pe \cite{daq} . This data flow rate is reduced by about a factor 1000 by a filter running on a PC farm on shore that looks for 5 L1 and a casual connection between L1 compatible with a light signal produced by a ultra-relativistic muon. The measured muon trigger rate is about 3 Hz in the 10 lines. In addition dedicated triggers have been developed for gamma-ray bursts (GRBs) and the Galactic Centre. A continuous record of up to 100 s stores temporarily data in memory that can be saved on disk in case of an external trigger from a network of satellites. The filter is necessary since sea water is an optical noisy environment due to bioluminescence and $\beta$ decay of $^{40}K$ producing electrons that annihilate and hence produce two photons above Cherenkov threshold. In 
the 10 line configuration we observed that during 80\% of the time the optical background rate is about 60 kHz.
Dedicated calibration runs are taken measuring the coincidence rate of 
 $^{40}K$ in the same storey and allowing to monitor OM efficiency at the level of 5\% accuracy. These efficiencies are used in the measurement of the rate of coincidences in the same storey versus depth as shown in Fig.~\ref{fig3} (on the left). The smooth  rate decrease is a measurement of muon energy losses in the detector and a demonstration that sea water is to a large extent a depth independent medium, unlike ice.  

The detection principle of a neutrino telescope (NT) is based on detection of Cherenkov light produced by ultra-relativistic muons at an angle of about $42^{\circ}$. Tracks are reconstructed using times of hits when light reaches PMTs and knowing the position of OMs. Energy can be reconstructed thanks to the charge measurement and the fact that muon energy losses above about 1 TeV are due to stochastic processes and the energy loss per unit length is proportional to energy. Muons are produced in charged current (CC) interactions of neutrinos that travel upward and have been filtered by the Earth out of the huge background of atmospheric muons. 
The neutrino cross section is extremely small ($\sim 10^{-34}$ cm$^2$ at $E_{\nu} \sim 30$ TeV)
hence the detectors have to cover large volumes to maximize the detection probability. Moreover, muons travel longer distances the higher their energy (eg. the average range of muons of 10 TeV is about 7.5 km), hence this technique is clearly optimized for best performances at very high energies. At 10 TeV the effective area of ANTARES for neutrinos is about 0.5 m$^2$ and at 1 PeV it is 10 m$^2$ averaging over all directions. Given the small cross section of neutrinos these numbers are very small compared to the geometrical dimension of these arrays. Notice also that the Earth begins to have a shadowing effect that depends on the nadir angle of neutrinos above 1 PeV due to the rise of the CC interaction cross section with energy. NTs are also sensitive to electron and tau neutrinos that are seen as more round sources of lights since cascades are small respect to the mutual separation between sensors. For this topology of events, nonetheless, the angular resolution is not as good as for muons which are then the best sample to point back to the direction of sources. 
 
NTs are located in sea or ice depths to guarantee a dark, transparent and well screened environment. Muons with energies above some TeV penetrate at a rate that at ANTARES depth is about 6 orders of magnitude larger than neutrinos. Given that NTs are detectors where photosensors have density limited by the need to cover large volumes, misreconstructed events can simulate upward-going neutrino-induced muons. This can happen especially for corner clipper events or large muon bundles where the time pattern can be confusing. The tracking capabilities can be also challenged in the presence of optical noise with baselines larger than the usual values around 60-100 kHz.  In order to demonstrate the ability to measure neutrinos, tracking algorithms must show that quality cut parameters allow to reconstruct a zenith angle distribution that shows their onset on top of the much higher background of atmospheric muons as shown in Fig.~\ref{fig3}(on the right). The plot is for 120.5 d of 5
line data and MC. This comparison shows a very good agreement in shape and a disagreement in normalization that is compatible with an uncertainty of 30\%,  mainly due to theoretical errors (cosmic ray composition and hadronic models), OM angular acceptance and attenuation length in sea water. 
Tracks are reconstructed using a linear fit algorithm that follows a hit selection based on time.  In Fig.~\ref{fig3} more than a line is required and at least 6 storeys and a quality cut on the reconstruction that accounts for hit amplitudes is applied. After this single cut, 1.29 atmospheric neutrinos per day (multi-line events) and 0.55/day (in the case of single line events) are selected, to be compared to an expectation of 1.22 from atmospheric neutrinos and 0.01 of misreconstructed atmospheric muons (0.62/day multi-line and 0.02/day single-line events). 
For the configuration of 10 strings 2 upward-going muons per day are selected involving more lines and 0.77/day single line events in 109 d.

\section{Searches for extraterrestrial neutrinos}	

The main purpose of NTs is to look for sources of high energy neutrinos on top of a large background due to atmospheric neutrinos. Searches for diffuse fluxes from sources distributed all over the ANTARES visible sky have to rely on the expectation that neutrinos accelerated in sources have a harder spectrum (close to $E^{-2}$ for $1^{st}$ order acceleration Fermi mechanism)
than atmospheric neutrinos. These are the result of cosmic ray interactions producing showers where the parent mesons (pions and kaons) decay or interact with probabilities that depend on column density of the atmosphere and energy (about $E^{-3.6}$ above 100 GeV). At energies larger than about 100 TeV the charm component coming from prompt decays of charmed particles begins to be important since it keeps the hard spectrum of primaries (about  $E^{-2.7}$ below the knee at about 4000 TeV for proton primaries).
These searches are more challenging than point-like source searches since this very high energy tail of the atmospheric neutrino spectrum is not well known given that accelerator measurement provide limited information. Moreover at these very high energies when the Earth is opaque to neutrinos the horizontal region provides the larger acceptance but even small errors in tracking can be critical for discriminating atmospheric muon events and muons from neutrinos. Finally the energy resolution of NTs is about a factor of 2, hence reconstructions of spectra are affected by large errors. 

Steady point-source searches on the other hand are very promising since they look for not only the energy spectrum signature but also for a clusterization of events on top of the large amount of atmospheric neutrinos. Recently more sophisticated methods than cluster searches with predefined angular windows have been developed. These show up to be more powerful by up to 50\% compared to standard binned methods \cite{braun,em}. The presence of a large background on top of a small directional signal allows to scramble times of events in declination bands creating equivalent samples of 
background events (null hypothesis). The signal+background hypothesis can be compared to the null hypothesis using various methods that use information that can distinguish signal respect to background.
Also  time signatures, such as in the case of GRBs, can help to discriminate atmospheric neutrinos and  residual atmospheric muon backgrounds.

\begin{figure}[pbt]
\begin{center}
\begin{tabular}{cc}
\psfig{file=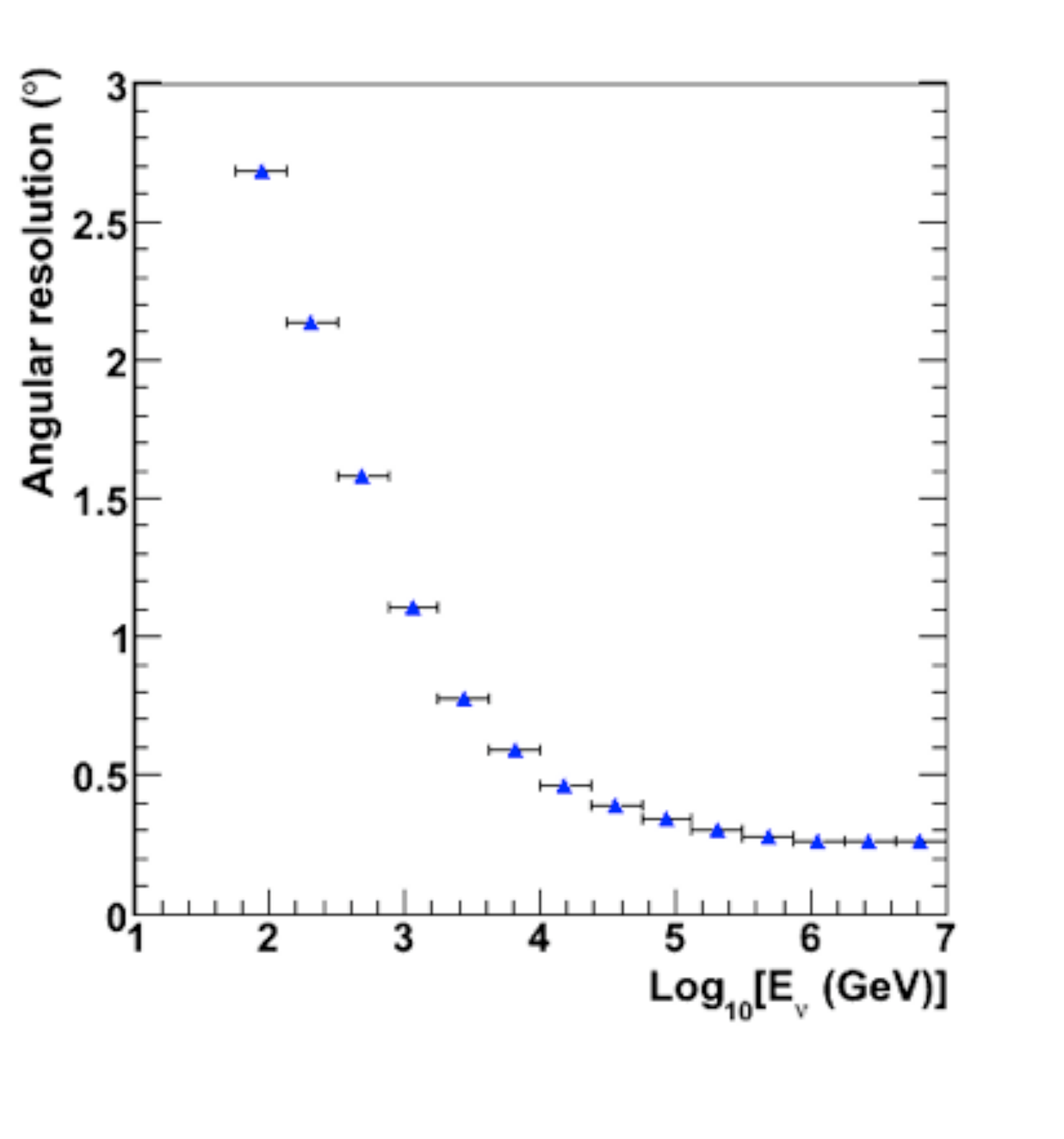, width=5.5cm,height=6.cm}&
\psfig{file=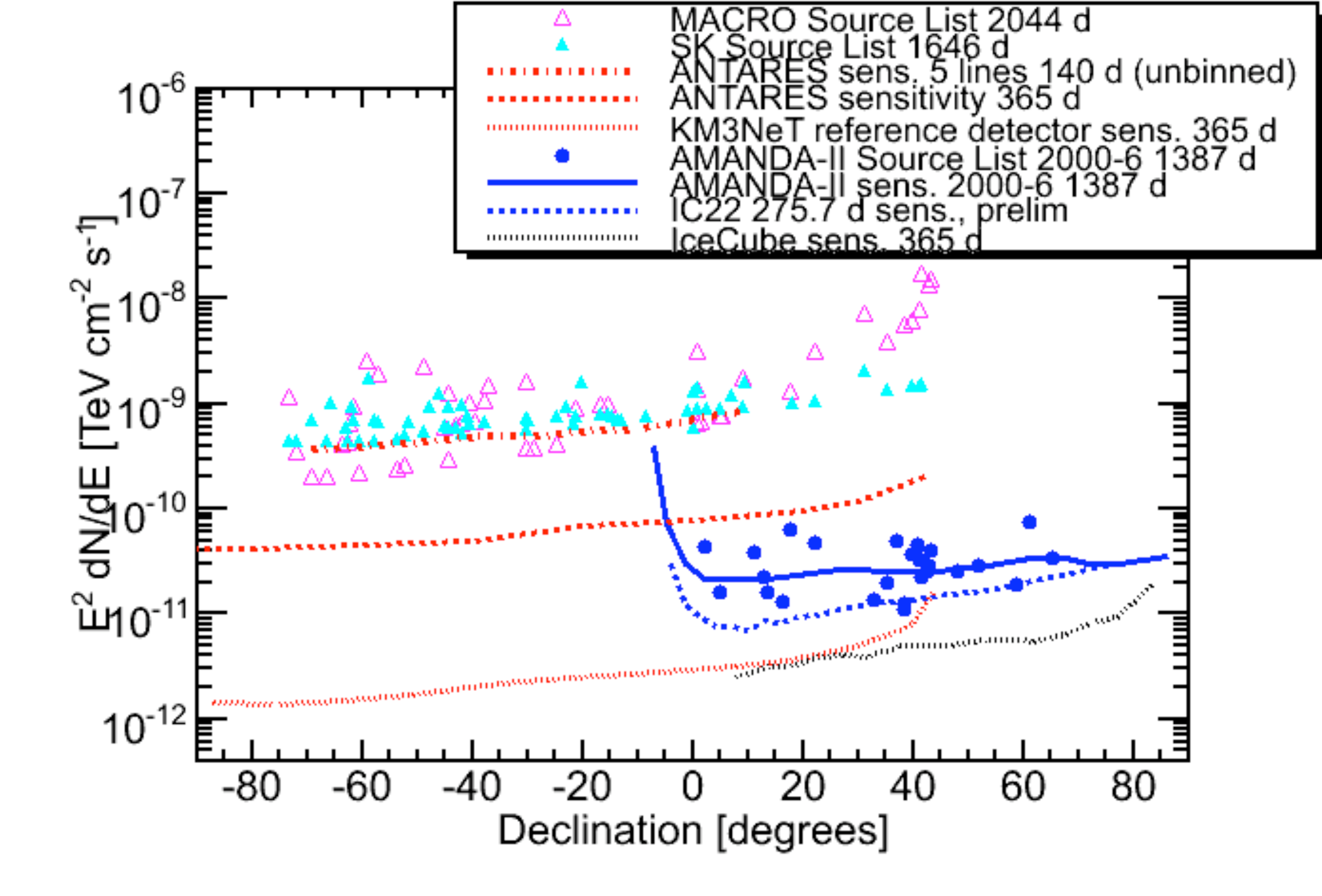, width=7.cm,height=6cm}
\end{tabular}
\end{center}
\caption{
On the left: Angular resolution (median of the distribution between the true neutrino direction and the reconstructed muon one) vs neutrino energy from simulations of the 5 lines detector for the quality cut on reconstruction. $\Lambda > -4.7$ is was chosen before unblinding data. On the right: sensitivity of 5 lines of ANTARES after 140 d (upper dashed-dotted line) to $E^{-2}$ fluxes of neutrinos vs declination. Also shown: the predicted result  
for 1 yr of the full ANTARES configuration (dashed line) and for a configuration of a possible km$^3$ detector in the Mediterranean (dotted line) \protect\cite{CDR}. Upper limits for catalogues of selected  sources are shown for Super-Kamiokande 
(full triangles) \protect\cite{sk} and MACRO (empty triangles) \protect\cite{macro}. Fort the opposite hemisphere the results of AMANDA-II
are shown (all-sky sensitivity - solid line - and limits for a catalogue of sources - full circles) \protect\cite{amanda7}, the sensitivity for 22 strings (dashed line) and 80 strings (dotted line) is also shown \protect\cite{now2008}.\label{fig4}}
\end{figure}

The 5-line data have been used to estimate the sensitivity and discovery potential for point like sources in the Southern hemisphere. At the level of the optimized cuts, the angular resolution for 5-lines is below $0.5^{\circ}$ above 10 TeV as shown in Fig.~\ref{fig4} (on the left).
The sensitivity for the all-sky search is shown in Fig.~\ref{fig4} (on the right). Though this is not the full configuration and reconstructions are still under development, the sensitivity for 140 d is already at the level of many years of detection of first generation experiments \cite{sk,macro}. The estimated result for the final configuration and a full year of data taking is also shown.

\begin{figure}[pb]
\begin{center}
\psfig{file=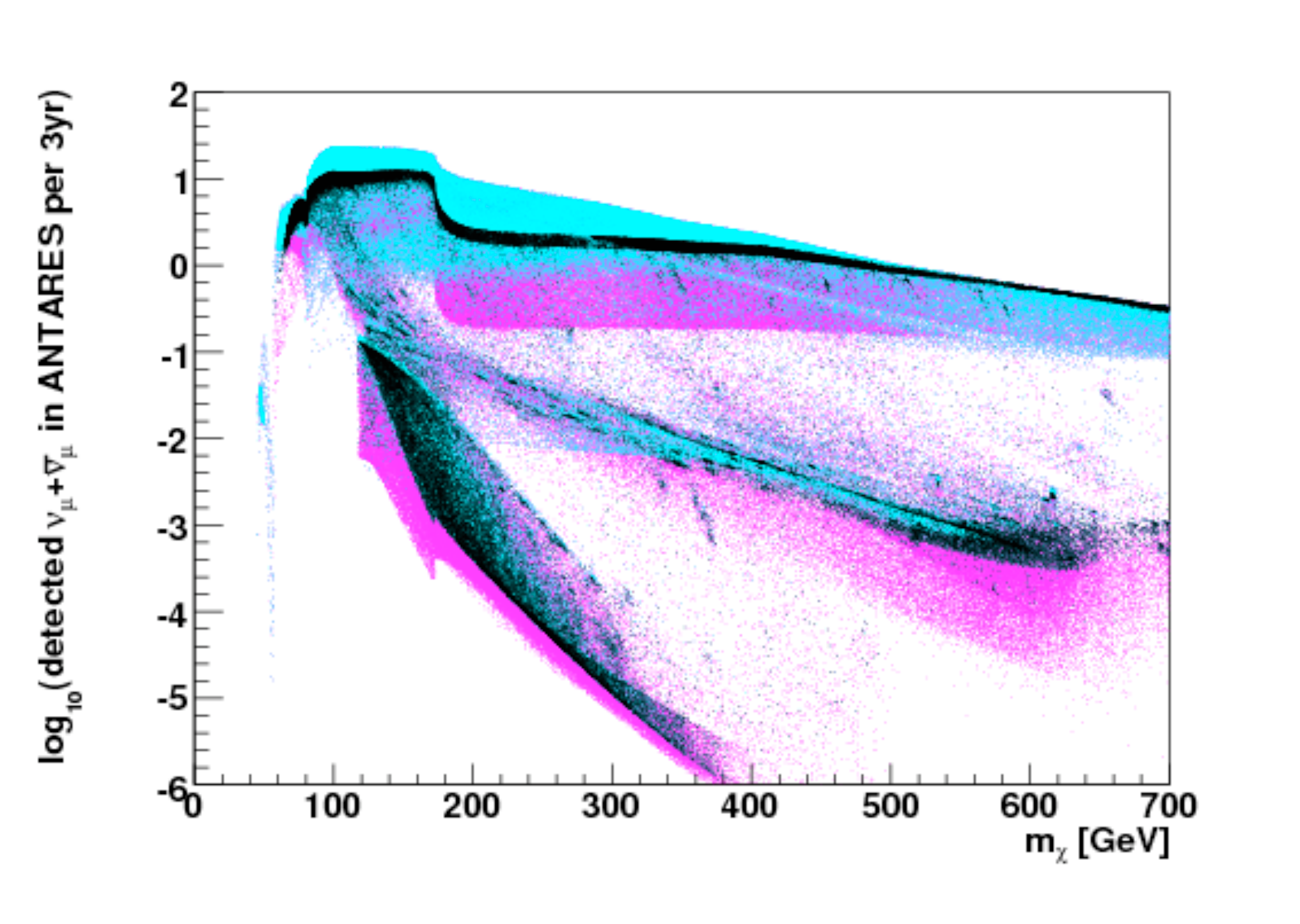, width=9.cm,height=8cm}
\end{center}
\caption{Number of events detectable by ANTARES in 3 yrs vs neutralino mass. Color codes refer to different relic densities (light blue $\Omega_{\chi} < 0.094$; black = within 2$\sigma$ of WMAP allowed range (0.094-0.129)\protect\cite{dm}; purple between 0.129 and 1.0.
\label{fig5}}
\end{figure}

Dark matter searches are also possible by looking for an excess of neutrinos from celestial bodies like the Sun or the Galactic Centre. Neutrinos with energy below TeV could be produced by the annihilation of supersymmetric dark matter particles like the neutralino, that would become gravitationally trapped in this bodies. In Fig.~\ref{fig5} we show the number of muon neutrino events that could be measured in 3 yrs by ANTARES in the direction of the Sun as a function the neutralino mass. Each of the point corresponds to a supersymmetric model.


\end{document}